\documentclass[manuscript]{acmart}

\usepackage{longtable}
\usepackage{tabularx}
\usepackage{array}
\usepackage{xcolor}
\usepackage{booktabs}
\usepackage{multirow}
\usepackage{makecell}

\renewcommand{\arraystretch}{1.5}

\newcolumntype{C}{>{\centering\arraybackslash}X}
\newcolumntype{Y}{>{\raggedright\arraybackslash}X}

\renewcommand\footnotetextcopyrightpermission[1]{}

\AtBeginDocument{%
  }

\begin{document}

\title{Co-Writing with AI: An Empirical Study of Diverse Academic Writing Workflows}
\renewcommand{\shorttitle}{Co-Writing with AI}

\author{Silvia Bodei}
\affiliation{%
  \institution{University College London}
  \city{London}
  \country{United Kingdom}}
\email{silvia.bodei.21@ucl.ac.uk}
\orcid{0009-0006-1997-1802}

\author{Duncan Brumby}
\affiliation{%
 \institution{University College London}
  \city{London}
  \country{United Kingdom}}
\email{d.brumby@ucl.ac.uk}
\orcid{0000-0003-2846-2592}

\author{Katie Fisher}
\affiliation{%
 \institution{University College London}
  \city{London}
  \country{United Kingdom}}
\email{katie.fisher@ucl.ac.uk}
\orcid{0009-0009-5552-3996}

\author{Jon Mella}
\affiliation{%
 \institution{University College London}
  \city{London}
  \country{United Kingdom}}
\email{jon.mella.21@ucl.ac.uk}
\orcid{0000-0002-1385-6727}

\renewcommand{\shortauthors}{Bodei et al.}

\begin{abstract}
Despite AI tools becoming increasingly embedded in academic practice, little is known about how university students integrate them into their writing processes. We examine how students engage with AI across different writing tasks, and how this engagement is shaped by individual factors including AI literacy, writing confidence, trust, authorship concerns, and motivation. Study~1 surveys 107 UK university students to map task-specific and co-occurring patterns of AI use across five writing stages (ideation, sourcing, planning, drafting, and reviewing) and their associations with individual factors. Study~2 complements this by exploring how these patterns can be assembled in practice, through interviews with 12 postgraduates reflecting on their established use of AI in assessed writing. Together, the studies suggest that AI integration is selective and heterogeneous, forming three recurring and value-oriented configurations: (1) early-stage (learning-oriented), where tools support exploration and understanding; (2) late-stage (quality-oriented), where tools support drafting and refinement; and (3) peripheral (productivity-oriented), where tools are used to reduce friction and sustain momentum across the process. We offer a workflow-level account of AI-supported academic writing, showing how students navigate competing priorities of learning, quality, productivity, and authorship, and how they evaluate and take responsibility for AI-generated outputs.
\end{abstract}

\ccsdesc[500]{Human-centered computing~Natural language interfaces}
\ccsdesc[500]{Applied computing~Education}

\keywords{Human-Computer Interaction (HCI), Generative AI, Academic Writing, Human-AI collaboration, Large Language Models (LLMs), Academic Integrity, Digital Tools in Education}

\maketitle

\noindent\textbf{Note.} This is the author's accepted manuscript of a paper accepted at CHIWORK 2026. The final version will appear in the ACM Digital Library.


\section{Introduction}

Writing is widely recognized as a foundational skill in knowledge work and a key site of competence development in formal education, demanding sustained effort, planning, synthesis, critical evaluation, and the development of a personal authorial voice \cite{elander_complex_2006}. These capacities are traditionally cultivated through repeated practice and formal assessment, but AI-powered writing tools have begun to reshape this landscape. Since the release of ChatGPT in late 2022, these systems have become widely used in academic practice, allowing users to generate fluent, contextually appropriate text from open-ended prompts. Adoption across student populations has been rapid, though large-scale studies also point to persistent gaps in AI literacy and an ongoing need for effective training and regulation \cite{freeman_student_2025, microsoft_ai_education_2025}.

The integration of AI into academic workflows raises critical questions about productivity, cognitive engagement, and the role of the writing process itself. While AI can enhance efficiency and writing quality, it may also encourage shallow engagement, reduce cognitive involvement, and foster over-reliance on automated systems. Evidence suggests that AI use may distort metacognitive judgments, leading users to overestimate their own performance \cite{fernandes_performance_2026}, and reshape critical thinking skills and practices \cite{LeeCHI2025}. Mixed human--AI workflows may also create difficulties around remembering what was written and correctly attributing its source \cite{ZindulkaCHI2026}, and there is a risk that users may simply fail to critically evaluate AI-generated outputs, allowing errors to go undetected or unchallenged \cite{gould_chattldr_2024}. These concerns have important implications for educational contexts, where students must retain cognitive control and responsibility while drawing on AI's capabilities.

In response to these developments, educational institutions have increasingly moved away from blanket prohibitions and toward fostering AI literacy and critical tool use among students \cite{lintner_systematic_2024}. Yet such guidance often treats "critical use" as a broadly uniform concept, emphasizing general principles such as awareness of limitations, ethical boundaries, and institutional policies. It therefore offers limited insight into how students should integrate AI into their writing under different purposes and constraints \cite{luo_jess_critical_2024, perkins_artificial_2024}.

Research has begun to examine student engagement with AI in more nuanced ways, identifying adaptive usage patterns and moderating individual factors such as AI literacy, writing confidence, trust, ethical stance, and authorial identity \cite{ou_academic_2024, chan_expectancy_2023, strzelecki_use_2024, tankelevitch_metacognitive_2024}. Few studies, however, examine how these individual differences translate into cohesive workflows across the writing process, or how patterns of AI use are organized across distinct tasks in ways that reflect broader goals of learning, output quality, productivity, and authorship.

To address this gap, we investigate how university students integrate AI tools across analytically defined writing stages --- ideation, sourcing, planning, drafting, and reviewing --- drawing on prior qualitative and mixed-methods research on human–AI writing \cite{nguyen_human_ai_2024, kim_exploring_2025, yao_qualitative_2025, shibani_untangling_2024}. Specifically, we address three research questions:

\begin{itemize}
\item RQ1: How do students integrate AI writing tools at different stages of the academic writing process?
\item RQ2: What individual factors are associated with AI use at each stage?
\item RQ3: How are these patterns assembled into coherent writing workflows?
\end{itemize}

We employ a mixed-methods design. Study~1 uses a survey of undergraduate and postgraduate students to map stage-specific and co-occurring patterns of AI use, examining their associations with individual factors such as AI literacy, writing confidence and various aspects of expected impact on writing output. The findings indicate that, although AI use is widespread, it is typically concentrated within a limited set of stages, with substantial variation across individuals. Co-occurrence analysis reveals three recurring configurations: early-stage use (ideation, sourcing, planning), late-stage use (drafting, reviewing), and peripheral use linking early and later stages. Each configuration is associated with distinct factors, including trust, perceived productivity, quality, and authorship, which become more salient at different points in the writing process.

Study~2 complements these findings through interviews with participants from a postgraduate program where AI use is permitted in an assistive role and supported through training, enabling more situated, open, and reflective accounts of AI integration. This approach addresses known limitations of self-report measures, particularly the tendency to underreport AI use due to social desirability bias and selective disclosure \cite{ling_underreporting_2026}. Prior work suggests that such underreporting is shaped not only by institutional ambiguity but also by emerging forms of "AI shaming," whereby AI use becomes negatively associated with competence or authorship \cite{sarkar_ai_2025}. Interviews provide a setting in which these tensions can be more openly articulated, particularly in contexts that permit AI use and support candid disclosure \cite{ruitenburg_what_2026}.

Within this framework, Study~2 offers an illustrative account of how stage-based patterns coalesce into cohesive workflows. Three configurations are observed: Learning-Oriented, where AI supports exploration and understanding; Quality-Oriented, where AI is used selectively to refine and improve text; and Productivity-Oriented, where AI reduces friction and sustains momentum across the process.

Together, the studies provide a workflow-level account of AI-supported academic writing, showing how students organize use around competing priorities of learning, quality, productivity, and authorship. This offers a basis for understanding variation in AI-supported writing practices, with implications for the design of task-aware writing systems and pedagogical approaches to critical AI literacy.

\section{Related Work}

\subsection{Emerging Challenges and Institutional Responses to AI in Academic Writing}

Academic writing constitutes a central and deeply contested domain of human–AI collaboration. Although AI use in written assessments is increasingly pervasive and difficult to regulate, higher education remains structurally dependent on written coursework as a scalable means of evaluating student performance and a key site for developing analytical reasoning, synthesis, and discipline-specific communication \cite{elander_complex_2006}. This tension raises critical questions about how AI use should be addressed within assessment practices without undermining their role in student learning \cite{elander_complex_2006}.

These concerns have been intensified by recent advances in generative AI. Earlier forms of AI writing support were largely limited to Automated Writing Evaluation (AWE) tools such as Grammarly or Criterion, which relied on rule-based systems to provide corrective feedback on pre-written text and targeted primarily surface-level features such as grammar and vocabulary. Generative AI systems such as ChatGPT, by contrast, employ large language models to produce fluent, context-aware text from minimal prompts, enabling use across a range of writing tasks beyond correcting completed text, such as finding sources, drafting, and substantial revision \cite{shibani_untangling_2024}. 

Reviews of institutional responses suggest that educational infrastructures have not kept pace with these developments. Existing approaches have been characterized along a continuum from prohibition to full integration, with restrictive policies predominating despite the practical difficulty of enforcing outright bans \cite{perkins_artificial_2024}. Much of the observed institutional guidance focuses on constraining specific forms of AI misuse, such as plagiarism, unacknowledged text generation, and hallucinated sources, rather than on how AI might be meaningfully embedded across the writing process, leaving the coordination of productive use largely unaddressed \cite{perkins_artificial_2024}.

\subsection{Critical AI Use and Early Adaptive Stage-Based Models of Writing}

In contrast to institutionally focused guidance, early empirical work examined student–AI collaboration in more adaptive and task-sensitive ways, moving away from static use/non-use models toward accounts of dynamic interaction with productive potential.

Pioneering laboratory studies provide an empirically grounded foundation for this shift. Using writing tasks designed to elicit natural AI interaction, Nguyen \cite{nguyen_hybrid_2024, nguyen_human_ai_2024} employed sequential log analyses to identify two distinct patterns of use: a goal-oriented, linear approach involving minimal system engagement, and an exploratory, iterative approach characterized by frequent reciprocal interaction. Follow-up expert evaluations and self-report measures found that iterative engagement was associated with higher-quality writing, greater satisfaction, and a stronger sense of retained authorship. Kim et al. \cite{kim_exploring_2025} expanded on this using think-aloud protocols, screen recordings, and interaction logs to compare high- and low-literacy users, finding that high-literacy students adopted a collaborative stance and flexibly adapted AI use across writing tasks, whereas low-literacy students exhibited limited interaction and minimal strategic adaptation. In follow-up interviews, students who engaged critically and tailored AI use to specific writing contexts reported improvements in perceived quality, motivation, and self-efficacy \cite{kim_students-generative_2025}. Interpreted through frameworks of self- and socially regulated learning \cite{jarvela_predicting_2023}, these findings suggest that students who engage iteratively and dialogically with AI demonstrate more adaptive and goal-directed writing strategies than those who passively accept AI-generated content, leading to improved writing outcomes.

Naturalistic qualitative studies further support this picture. Yao et al. \cite{yao_qualitative_2025}, in interviews with high-performing postgraduates, identified defined adaptive metacognitive strategies --- planning, monitoring, evaluating, information management, and debugging --- that structured differentiated AI integration across writing tasks. Parker et al. \cite{parker_negotiating_2024} similarly found, through focus groups with doctoral students, that successful AI integration involves continuous negotiation between tool engagement and writer agency, producing heterogeneous interaction patterns rather than uniform workflows. 

Shibani et al. \cite{shibani_untangling_2024} offer a structured theoretical account of these dynamics in the Critical Interaction with AI in Writing (CIAW) framework. Drawing on longitudinal data combining writing samples, interaction logs, and reflective prompts, the framework organizes writing practices across three activity domains --- ideation and planning, information seeking and evaluation, and presentation and writing --- and conceptualizes engagement through three analytically distinct modes: absent, shallow, and deep use. Absent use refers to writing activity in which AI is not engaged. Shallow use involves limited critical mediation, typically characterized by one-off prompting, minimal evaluation or revision, and direct incorporation of AI outputs. Deep use, by contrast, is defined by sustained, iterative, and dialogic interaction, involving cycles of prompting, evaluation, modification, and reinterpretation through which content is jointly constructed rather than passively adopted.

Notably, while this body of work reflects a clear shift toward dynamic and adaptive models of student–AI collaboration, much of it emerged during the early phase of AI adoption, when institutional responses remained largely restrictive and student familiarity with AI systems was limited \cite{arowosegbe_perception_2024, ou_academic_2024}. Consequently, definitions of critical use have often remained anchored in foundational competencies --- recognizing system limitations, evaluating outputs, and avoiding inaccuracies --- while offering limited insight into how students organize AI use across complete writing workflows. As AI becomes more deeply embedded in academic practice, students are likely to develop increasingly personalized and heterogeneous configurations of use \cite{arowosegbe_perception_2024}, yet existing frameworks are poorly equipped to account for this contextual and individual variability \cite{ou_academic_2024, arowosegbe_perception_2024}.

\subsection{Individual Factors Shaping Stage-Specific AI Integration in Academic Writing}

While empirical work directly examining workflow variation across writing stages remains limited, converging evidence from higher education students, university-recruited samples, and knowledge workers suggests that AI adoption is consistently shaped by individual dispositions, task demands, and perceived value. Three strands of this work are particularly relevant.

One strand, drawing primarily on higher education student populations, examines how perceptions of task fit, motivation, and expected outcomes shape engagement with AI tools \cite{malik_exploring_2023, strzelecki_use_2024}. Findings from this work indicate that AI engagement is not purely instrumental but involves normative assessments of the appropriateness of assistance for particular writing tasks. Malik et al. \cite{malik_exploring_2023}, for instance, found that while perceived usefulness predicts engagement, ethical stance and the desire to maintain authorial identity can override efficiency considerations, suggesting that adoption reflects value-laden trade-offs rather than simple cost-benefit calculations.

A second strand, grounded in social constructivist perspectives and focused predominantly on student contexts, conceptualizes AI as a cognitive and pedagogical partner within learning processes. Drawing on Vygotsky's Zone of Proximal Development \cite{vygotsky_mind_1978}, this work shows that AI adoption is often motivated by perceived learning benefits but constrained by concerns about dependency, integrity, and authorship \cite{boillos_student_2025, song_enhancing_2023}. Classroom interventions have demonstrated improvements in writing quality and motivation, yet have also surfaced concerns about ownership and over-reliance \cite{song_enhancing_2023}, while Boillos and Idoiaga \cite{boillos_student_2025} found that students valued AI's efficiency and quality gains but feared skill atrophy and loss of integrity. Together, these findings suggest that AI is most readily adopted when it scaffolds learning without undermining independent competence, reinforcing the broader pattern of value-based trade-offs identified across this literature.

A third strand, drawing on broader populations including knowledge workers and university-recruited samples, conceptualizes AI as an external cognitive resource that redistributes effort across tasks. Dhillon et al. \cite{dhillon_shaping_2024}, working with university-recruited participants on writing tasks, compared different levels of AI scaffolding and found that while increased support reduced cognitive load and improved writing quality, it also lowered motivation and perceived authorship. Fernandes et al. \cite{fernandes_performance_2026} similarly show that performance gains can occur alongside reduced metacognitive awareness, suggesting a shift in how cognitive resources are allocated. Tankelevitch et al. \cite{tankelevitch_metacognitive_2024}, drawing on knowledge work contexts, further argue that while AI assistance eases task demands, it may alter metacognitive engagement in ways that affect depth of involvement. Complementing this, Lee et al. \cite{LeeCHI2025} find that higher confidence in AI is associated with reduced critical thinking effort, with cognitive activity shifting away from generative reasoning and toward evaluative and coordinative functions such as verifying outputs, integrating responses, and overseeing task execution. Zindulka et al. \cite{ZindulkaCHI2026} further show that AI use can affect perceptions of authorial self-attribution, with participants in human–AI collaboration conditions becoming less able than control groups to identify the source of co-created ideas after a delay. Recent work on collaborative document editing finds that even when AI is integrated as a shared resource across a team, users incorporate it into existing norms of authorship and coordination rather than treating it as an autonomous contributor \cite{LehmannCHI2026}, suggesting that individual values and boundaries around authorship remain central to AI integration regardless of the scale or setting of the writing task. Collectively, findings grounded in cognitive approaches suggest that the benefits of AI depend not only on reducing effort, but on how effort is allocated and whether users retain sufficient engagement to evaluate and take ownership of their work. Yet, despite the richness of these accounts, comparable evidence for how students specifically negotiate these boundaries across the stages of academic writing remains scarce. 

\subsection{Toward a Workflow-Level Account of AI Use}

Across the literature, critical AI use emerges not as a uniform skill but as a situated judgment shaped by individual dispositions, including AI literacy, confidence, trust, and authorial identity, and by anticipated trade-offs in text quality, ownership, voice, motivation, and learning \cite{dhillon_shaping_2024, malik_exploring_2023}. Rather than converging on a single pattern of use, these factors likely lead students to allocate responsibility to AI selectively across the writing process, with different stages becoming sites of delegation or resistance depending on what is at stake \cite{arowosegbe_perception_2024, shibani_untangling_2024}. 

Despite these insights, few studies explicitly connect individual-level factors to where within the writing process AI is used. Academic writing is still commonly treated as a relatively uniform activity, obscuring how engagement varies across distinct stages and tasks. While broader workflow-oriented analyses exist \cite{kim_exploring_2025, kim_students-generative_2025, nguyen_human_ai_2024, nguyen_hybrid_2024, parker_negotiating_2024, yao_qualitative_2025}, they typically conceptualize critical use as a single construct focused on awareness of tool limitations and functionality, overlooking how individual differences shape stage-specific integration or how patterns of use interact across tasks within coherent workflows \cite{arowosegbe_perception_2024, ou_academic_2024}.

The present study addresses this gap by examining how students integrate AI tools across stages of academic writing (RQ1), which individual factors are associated with AI use at each stage (RQ2), and how these patterns combine in the workflow designs of critically engaged students (RQ3). Study 1 investigates RQ1 and RQ2 through a quantitative survey mapping AI use across five writing stages and its associations with individual factors. Study 2 addresses RQ3 through in-depth interviews, examining how these patterns are articulated, enacted, and assembled into coherent, end-to-end writing workflows. This research was conducted as part of a master’s dissertation \cite{Bodei_2025}, on which the present paper builds.

\section{Study 1: Quantitative Survey of Stage-Specific AI Use and Influencing Factors}

\subsection{Method}

\subsubsection{Participants}

A total of 186 students initiated the survey. Participants who completed less than 80\% of the instrument were excluded to ensure adequate coverage and reduce bias from partial responses, consistent with established survey design principles \cite{dillman_mail_2007}. This yielded an analytic sample of 107 students from UK institutions. Recruitment used convenience sampling across both online and offline channels: 70 participants were recruited via Prolific and compensated at the institutional rate (£3), while 37 participated voluntarily through personal networks. Eligibility required current enrollment at a UK university, with data collection restricted to the UK to maintain GDPR compliance. The final sample comprised undergraduates ($n$ = 52), master’s students ($n$ = 48), and doctoral students ($n$ = 7) from diverse academic disciplines. To capture the full range of user experience and integration patterns, participants were not pre-screened for prior AI use. Institutional ethical approval was obtained prior to data collection.

\subsubsection{Materials and Procedure}

The study employed an \textit{ad hoc} survey instrument with embedded consent to collect background data on AI use in academic writing, stage-specific usage patterns, and ratings of influencing factors identified in prior literature. The instrument integrated validated scales where available and included targeted items aligned with the study objectives.

Instrument development proceeded iteratively. An initial pilot version, tested with 11 participants, included open-ended items on perceived advantages, disadvantages, and motivations for AI use at each stage. These items were excluded from the final survey, as responses were typically brief and repetitive, contributing to participant fatigue without yielding additional insight. The final survey was administered online via Qualtrics, required approximately 20 minutes to complete, and presented items in a fixed order. The following sections detail each component of the survey.

\textit{Background data}: After reviewing the information sheet and providing consent, participants reported demographic details (university, course, level of study, and prior education) to verify eligibility and characterize the sample. They then indicated the frequency and types of assessments for which they had used AI, their most commonly used tools, and overall usage frequency. These questions contextualized later responses and enabled comparison with prior large-scale descriptive surveys of student AI use \cite{freeman_student_2025}.

To further contextualize responses, participants rated their department’s stance on AI in assessment using Perkins et al.’s Artificial Intelligence Assessment Scale \cite{perkins_artificial_2024}, a single-item measure classifying institutional positions on a 5-point scale from full prohibition (1) to full acceptance and encouragement (5). Participants also completed three \textit{ad hoc} 5-point Likert items assessing their comfort with disclosing AI use to peers, in written work, and to lecturers (1 = very uncomfortable, 5 = very comfortable), capturing potential reticence in reporting AI practices despite the anonymity of the survey.

\textit{Writing confidence and AI literacy}: Participants next rated their academic writing confidence and AI literacy, both identified as moderators of AI integration in prior work. Writing confidence was measured using three 5-point Likert items (1 = lowest confidence, 5 = highest confidence) assessing perceived writing ability, typical feedback received, and ability relative to peers, following recommendations to include feedback history alongside self-perception to reduce self-report bias \cite{lee_design_2024}. AI literacy was measured using the 7-item \emph{Knowledge of Generative AI} subscale from Chan and Zhou \cite{chan_expectancy_2023}, a validated 5-point Likert instrument (1 = lowest literacy, 5 = highest literacy) that is both robust and concise for assessing AI literacy in higher education \cite{lintner_systematic_2024}.

\textit{Stage-specific use}: To capture workflow-level patterns, participants first indicated the stages of the academic writing process in which they typically used AI: ideation, sourcing, planning, drafting, reviewing, or “I do not use AI at all.” The checklist was adapted from the Critical Interaction with AI in Writing (CIAW) framework \cite{shibani_untangling_2024}, which conceptualizes critical engagement across five dimensions. Three task-based dimensions --- Planning and Ideation, Information Seeking and Evaluation, and Writing and Presentation --- align with writing stages (ideation and planning, sourcing, and drafting and reviewing). Two broader dimensions, Reflection on AI-Assisted Learning and Conversational Engagement, capture metacognitive and dialogic processes spanning all stages. Although not included as checklist items, these dimensions informed the interpretation of reflective engagement across stages.

For each stage, participants completed a checklist of typical AI uses representing deep, shallow, or no engagement (“I do not use AI at this stage”). Each checklist contained realistic task examples derived from the CIAW framework, and participants selected all options that reflected their typical use. Within CIAW, shallow use denotes dependent engagement, where outputs are accepted with minimal evaluation, whereas deep use involves iterative refinement, critical review, and selective integration of AI suggestions. These distinctions were operationalized as stage-specific items reflecting authentic student writing practices.

For example, in ideation, shallow use included “using AI to generate general definitions or explanations of the assignment topic,” while deep use involved “iteratively discussing and refining potential topics with AI tools.” In sourcing, shallow use included “asking AI for a list of ‘top studies’ or a literature review,” whereas deep use involved “requesting information from AI and manually checking the reliability of sources.” In planning, shallow use reflected “asking AI for a standard essay structure,” while deep use included “using AI to identify gaps or redundancies in a self-drafted plan.” In drafting, shallow use involved “prompting AI to write a paragraph or section,” whereas deep use included “collaborating with AI to refine phrasing and argumentation through multiple iterations.” Finally, in reviewing, shallow use included “using AI to check grammar or spelling and pasting the output directly into my draft,” while deep use involved “asking AI to identify unclear or underdeveloped arguments and reflecting on how to improve them through targeted revisions.”

\textit{Final ad hoc factor ratings}: In the final section, participants rated a set of factors previously identified as influencing AI use in academic writing. These included individual characteristics (e.g., trust in AI, ethical stance, authorial identity) and perceived impacts on text and process (e.g., text quality, ownership, creative voice, productivity, motivation, learning outcomes). In the absence of concise validated scales suitable for this context, each factor was measured using a single \textit{ad hoc} statement rated on a 5-point Likert scale (1 = substantially negative, 5 = substantially positive). Items were designed to capture both the direction and strength of AI’s perceived influence, providing an intuitive and efficient measure of participants’ evaluations. Informal feedback from pilot participants ($n$ = 11) indicated that the items were clear and easy to interpret, supporting their use in the final instrument.

Example statements included “How do you feel AI affects the overall quality of your work?” (1 = makes it substantially worse, 5 = substantially improves it), “How much does including AI affect your feelings of authorship and ownership over your work?”, and “To what extent do you trust AI in your academic writing?”. While this approach involves trade-offs in reliability and construct validity, similar single-item measures have been used in recent studies of AI use for constructs such as trust, ethics, and authorship, particularly in contexts requiring efficient assessment across multiple dimensions \cite{malik_exploring_2023, dhillon_shaping_2024}. This design enabled the inclusion of a broad range of constructs within a single instrument, supporting exploratory analysis of how perceptual factors relate to AI use across writing stages while minimizing participant burden.

\subsubsection{Data Analysis}

Quantitative analyses were conducted in R. The analyses provide an exploratory mapping of how AI use varies across writing stages and how it relates to individual characteristics, rather than supporting confirmatory hypothesis testing or complex multivariate modeling. This strategy reflects the absence of directional hypotheses, the use of single-item measures for several constructs, and sample size and power constraints, which limit the robustness of more complex statistical approaches.

First, background measures (prior AI use, departmental stance, and disclosure comfort) were summarized using descriptive statistics, alongside AI literacy and academic writing confidence, to contextualize subsequent analyses.

Second, habitual AI use across writing stages was examined descriptively to characterize workflow patterns. Proportions of students reporting AI use at each stage were calculated, along with the average number of stages used per participant and the most frequent stage combinations (e.g., ideation–planning–drafting). Among participants reporting AI use at a given stage, the proportion endorsing each shallow or deep use item was computed to capture variation in engagement styles within and across stages.

Third, comparative analyses examined differences in individual characteristics between users and non-users at each writing stage. Mean scores on academic writing confidence, AI literacy, and the \textit{ad hoc} factors were compared between groups using pairwise tests. An \textit{a priori} power analysis indicated that approximately 64 observations per group (i.e., users vs.\ non-users within a given stage) would be required to detect medium-sized effects (Cohen’s $d \approx 0.5$) with 80\% power \cite{cohen_power_1992}. As participants were classified separately at each stage based on reported AI use, group sizes varied and did not consistently meet this threshold. Accordingly, these analyses are treated as exploratory and focus on identifying broad patterns of association rather than supporting more complex techniques requiring greater power (e.g., clustering or multivariate analyses). Welch’s independent-samples t-tests were used for their robustness to violations of normality and homogeneity of variance in smaller samples.

As the survey required page-by-page completion, missing data arose only when participants discontinued before the final factor ratings. Beyond excluding responses with less than 80\% completion, an available-case approach was adopted for the factor analyses, retaining participants with valid responses for each factor and excluding them only from analyses involving missing data (pairwise deletion). This approach preserves statistical power and is appropriate under a Missing at Random (MAR) assumption, as missingness is primarily attributable to survey attrition rather than selective item non-response \cite{little_statistical_2019}.

\subsection{Results}

\subsubsection{Background Data}

The majority of students (75\%) reported habitual use of AI tools in written assessments, while 23\% reported no use and 2\% were unsure. The most common applications were essays (54\% of students), coursework assignments (50\%), reports (45\%), and dissertation or thesis writing (35\%). ChatGPT dominated overall tool use, reported by 65\% of the sample, followed by Gemini (7\%), DeepSeek (6\%), Claude (2\%), and Perplexity (2\%). 

In terms of frequency, 33\% of students reported using AI multiple times per week and 28\% daily, while smaller proportions used it weekly (18\%), less than weekly (15\%), or not at all (6\%). Regarding departmental policies, 49\% indicated that AI is permitted for idea generation or brainstorming, whereas 21\% reported complete prohibition. Sixteen percent reported that AI-assisted editing is allowed for refining text but not generating original content, 8\% indicated that limited sections of assessments may include AI use, and only 6\% reported full encouragement of AI collaboration.

\begin{table}[t]
\centering
\footnotesize
\setlength{\tabcolsep}{4pt} 
\caption{Conditional co-occurrence of AI use across writing stages. Rows indicate the conditioning stage (i.e., the subset of students who used AI at that stage); columns indicate the stage for which AI use is reported. Each cell shows the percentage of students who reported using AI at both the row and column stages.}
\label{tab:cooccurrence}
\begin{tabularx}{\columnwidth}{l*{5}{>{\centering\arraybackslash}X}}
\toprule
 & \textbf{Ideation} & \textbf{Sourcing} & \textbf{Planning} & \textbf{Drafting} & \textbf{Reviewing} \\
\midrule
\textbf{Ideation}   & --  & 49\% & 52\% & 42\% & 48\% \\
\textbf{Sourcing}   & 66\% & --  & 52\% & 46\% & 48\% \\
\textbf{Planning}   & 63\% & 47\% & --  & 55\% & 57\% \\
\textbf{Drafting}   & 53\% & 43\% & 57\% & --  & 60\% \\
\textbf{Reviewing}  & 61\% & 46\% & 61\% & 61\% & --  \\
\bottomrule
\end{tabularx}
\end{table}

Students’ comfort in disclosing AI use was moderate overall ($M$ = 3.15, $SD$ = 1.20), highest when disclosing to peers ($M$ = 3.47, $SD$ = 1.36), followed by disclosure in written work ($M$ = 3.21, $SD$ = 1.33), and lowest when disclosing to lecturers ($M$ = 2.77, $SD$ = 1.43). AI literacy scores were relatively high ($M$ = 3.70, $SD$ = 0.99), as were academic writing confidence scores ($M$ = 3.73, $SD$ = 0.61).

\subsubsection{Stage-Specific Use}

AI use was relatively evenly distributed across writing stages, with no single stage emerging as the dominant point of reliance. The stage most frequently supported by AI was ideation (55\% of participants), followed by planning (46\%), drafting (44\%), reviewing (43\%), and sourcing (41\%). Notably, 9\% of participants reported not using AI at any stage.

To assess the breadth of integration, we calculated the number of stages in which each participant reported habitual AI use, assigning zero for participants who did not use AI at all. The average number of stages with reported AI use was $M$ = 2.38, $SD$ = 1.26. Most students reported using AI in one (28\%) or two (29\%) stages, while fewer used it in three (25\%), four (8\%), or all five stages (9\%). This suggests that students generally concentrate AI use on a limited subset of the writing process, rather than uniformly applying it across all stages.

To examine how students combine AI use across stages, we analyzed conditional co-occurrence patterns. Table~\ref{tab:cooccurrence} reports, for each row stage, the percentage of participants who reported using AI at that stage and also reported use in the corresponding column stage. For example, 49\% of participants who used AI for ideation also reported using it for sourcing. This representation highlights how use at one stage tends to co-occur with use at others. 

Patterns in the relative magnitude and concentration of these co-occurrence values, examined through descriptive analysis of stage-level associations, suggest three recurring configurations of AI integration across the writing process: (1) early-stage use centered on ideation, sourcing, and planning; (2) late-stage use centered on drafting and reviewing; and (3) cross-stage use linking early and later stages. Below, we characterize each configuration in relation to the observed stage-level patterns.

\paragraph{Early-stage cluster (ideation--sourcing--planning).}
Students in this cluster primarily used AI to support sensemaking, topic development, and structural preparation. In ideation, common activities included generating general explanations of the assignment topic (54\%, shallow), generating and selecting ideas (37\%, shallow), and comparing or refining ideas relative to project priorities (38\%, deep).

In sourcing, dominant uses included asking AI to summarize articles (61\%, shallow) or locate sources for specific claims (46\%, shallow). A smaller subset of students requested information and then independently verified sources (38\%, deep). These activities mirror the strong co-occurrence between ideation, sourcing, and planning observed in Table~\ref{tab:cooccurrence}.

\begin{table}[p]
\centering
\footnotesize
\renewcommand{\arraystretch}{1.05}
\setlength{\tabcolsep}{4pt}

\caption{Comparison of AI users and non-users across writing stages and individual factors. Values show group means ($M$, $SD$; $n$) for each group. All scales range from 1--5, with higher values indicating more favorable evaluations. Test statistics are reported as Welch’s $t$ (df). Significance is indicated with $^{*}p < .05$, $^{**}p < .01$, $^{***}p < .001$.}
\label{tab:factors_long}
\begin{tabularx}{\textwidth}{p{1.9cm}Y>{\centering\arraybackslash}p{2.7cm}>{\centering\arraybackslash}p{2.7cm}>{\centering\arraybackslash}p{1.9cm}}
\toprule
\textbf{Stage} & \textbf{Factor} &
\makecell[c]{\textbf{Users} \\ \small $M$ ($SD$); $n$} &
\makecell[c]{\textbf{Non-users} \\ \small $M$ ($SD$); $n$} &
\textbf{$t$ (df)} \\
\midrule

\textbf{Ideation}
& Academic writing confidence & 3.70 (0.64); 59 & 3.76 (0.58); 48 & 0.54 (103.6) \\
& AI literacy & 3.53 (1.06); 59 & 3.91 (0.89); 48 & 2.02 (104.9)$^{*}$ \\
& Authorial identity & 4.12 (0.93); 51 & 4.16 (1.25); 43 & 0.20 (76.3) \\
& Ethical favorability of AI use & 3.22 (0.94); 51 & 2.56 (1.10); 43 & -3.08 (83.5)$^{**}$ \\
& Impact on learning & 3.67 (1.05); 51 & 2.91 (1.25); 43 & -3.15 (82.4)$^{**}$ \\
& Impact on motivation/enjoyment & 3.61 (0.83); 51 & 3.16 (1.00); 43 & -2.33 (81.7)$^{*}$ \\
& Impact on personal style/voice & 3.29 (0.99); 51 & 2.70 (1.12); 43 & -2.71 (84.3)$^{**}$ \\
& Impact on productivity & 4.00 (0.75); 51 & 3.49 (1.05); 43 & -2.66 (74.0)$^{**}$ \\
& Impact on text ownership & 3.10 (0.90); 51 & 2.28 (1.01); 43 & -4.12 (85.1)$^{***}$ \\
& Impact on writing quality & 3.96 (0.85); 51 & 3.53 (1.10); 43 & -2.07 (78.2)$^{*}$ \\
& Trust in AI & 3.31 (1.05); 51 & 2.51 (1.08); 43 & -3.64 (88.5)$^{***}$ \\

\midrule

\textbf{Sourcing}
& Academic writing confidence & 3.70 (0.65); 44 & 3.75 (0.58); 63 & 0.44 (85.4) \\
& AI literacy & 3.48 (1.08); 44 & 3.64 (0.92); 63 & -1.30 (83.0) \\
& Authorial identity & 4.16 (1.00); 38 & 4.12 (1.15); 56 & -0.15 (86.2) \\
& Ethical favorability of AI use & 3.03 (1.05); 38 & 2.84 (1.07); 56 & -0.84 (80.7) \\
& Impact on learning & 3.58 (1.13); 38 & 3.14 (1.23); 56 & -1.77 (83.8) \\
& Impact on motivation/enjoyment & 3.58 (0.89); 38 & 3.30 (0.95); 56 & -1.30 (82.9) \\
& Impact on personal style/voice & 2.95 (0.93); 38 & 3.07 (1.19); 56 & 0.57 (90.1) \\
& Impact on productivity & 3.87 (0.93); 38 & 3.70 (0.93); 56 & 0.88 (79.5) \\
& Impact on text ownership & 2.79 (1.04); 38 & 2.68 (1.03); 56 & -0.51 (78.8) \\
& Impact on writing quality & 3.95 (0.93); 38 & 3.64 (1.02); 56 & -1.50 (84.2) \\
& Trust in AI & 3.08 (1.08); 38 & 3.30 (1.17); 56 & -0.46 (83.8) \\

\midrule

\textbf{Planning}
& Academic writing confidence & 3.72 (0.61); 49 & 3.74 (0.61); 58 & 0.12 (101.9) \\
& AI literacy & 3.66 (0.96); 49 & 3.63 (1.04); 58 & -0.19 (104.1) \\
& Authorial identity & 3.89 (1.17); 44 & 4.36 (0.96); 50 & 2.13 (83.7)$^{*}$ \\
& Ethical favorability of AI use & 3.05 (0.99); 44 & 2.80 (1.12); 50 & -1.13 (92.0) \\
& Impact on learning & 3.68 (0.96); 44 & 3.00 (1.31); 50 & -2.90 (89.2)$^{**}$ \\
& Impact on motivation/enjoyment & 3.52 (0.82); 44 & 3.30 (1.02); 50 & 1.18 (91.4) \\
& Impact on personal style/voice & 3.05 (0.96); 44 & 3.00 (1.20); 50 & -0.20 (91.3) \\
& Impact on productivity & 4.18 (0.58); 44 & 3.40 (1.03); 50 & -4.60 (79.1)$^{***}$ \\
& Impact on text ownership & 2.84 (1.01); 44 & 2.62 (1.05); 50 & -1.04 (91.2) \\
& Impact on writing quality & 4.11 (0.75); 44 & 3.46 (1.07); 50 & -3.45 (87.9)$^{***}$ \\
& Trust in AI & 3.16 (1.01); 44 & 3.04 (1.20); 50 & -0.84 (91.8) \\

\midrule

\textbf{Drafting}
& Academic writing confidence & 3.72 (0.64); 47 & 3.74 (0.59); 60 & 0.19 (94.7) \\
& AI literacy & 3.49 (1.03); 47 & 3.86 (0.95); 60 & 1.89 (95.3) \\
& Authorial identity & 3.73 (1.23); 41 & 4.45 (0.85); 53 & 3.22 (67.8)$^{**}$ \\
& Ethical favorability of AI use & 3.15 (1.04); 41 & 2.74 (1.06); 53 & -1.88 (87.0) \\
& Impact on learning & 3.46 (1.03); 41 & 3.21 (1.32); 53 & -1.06 (92.0) \\
& Impact on motivation/enjoyment & 3.46 (0.87); 41 & 3.36 (0.98); 53 & -0.22 (90.3) \\
& Impact on personal style/voice & 3.07 (1.10); 41 & 2.98 (1.08); 53 & -0.40 (85.3) \\
& Impact on productivity & 3.95 (0.86); 41 & 3.62 (0.97); 53 & -1.74 (90.0) \\
& Impact on text ownership & 3.00 (0.95); 41 & 2.51 (1.05); 53 & -2.37 (89.7)$^{*}$ \\
& Impact on writing quality & 4.00 (0.77); 41 & 3.58 (1.10); 53 & -2.15 (91.3)$^{*}$ \\
& Trust in AI & 3.29 (0.90); 41 & 2.68 (1.22); 53 & -2.80 (91.8)$^{**}$ \\

\midrule

\textbf{Reviewing}
& Academic writing confidence & 3.71 (0.60); 46 & 3.74 (0.62); 61 & 0.32 (99.2) \\
& AI literacy & 3.64 (1.06); 46 & 3.74 (0.96); 61 & 0.53 (91.5) \\
& Authorial identity & 3.77 (1.27); 39 & 4.40 (0.85); 55 & 2.71 (61.8)$^{**}$ \\
& Ethical favorability of AI use & 3.13 (1.06); 39 & 2.76 (1.05); 55 & -1.65 (81.9) \\
& Impact on learning & 3.44 (0.97); 39 & 3.24 (1.35); 55 & -0.95 (90.5) \\
& Impact on motivation/enjoyment & 3.59 (0.82); 39 & 3.27 (0.99); 55 & -1.70 (89.8) \\
& Impact on personal style/voice & 3.05 (1.07); 39 & 3.00 (1.11); 55 & -0.23 (83.4) \\
& Impact on productivity & 4.13 (0.73); 39 & 3.51 (0.98); 55 & -3.51 (91.7)$^{***}$ \\
& Impact on text ownership & 2.97 (0.96); 39 & 2.55 (1.05); 55 & -2.05 (86.2)$^{*}$ \\
& Impact on writing quality & 4.28 (0.60); 39 & 3.40 (1.05); 55 & -5.15 (88.8)$^{***}$ \\
& Trust in AI & 3.15 (0.93); 39 & 2.80 (1.24); 55 & -1.58 (91.6) \\

\bottomrule
\end{tabularx}
\end{table}

Planning similarly combined shallow and deep engagement, including asking AI for an outline to initiate drafting (54\%, shallow) and using AI to order or reorganize points (46\%, deep). Together, these patterns suggest that early-stage AI use is oriented toward exploration, sensemaking, and reducing the cognitive cost of getting started, rather than producing final text.

\paragraph{Late-stage cluster (drafting--reviewing).}
Students in this cluster concentrated AI use on text production and refinement. During drafting, the most common strategy was providing a rough paragraph and requesting improvements to flow or argument strength, then selectively incorporating suggestions (49\%, deep), alongside rephrasing awkward sentences (41\%, shallow) and converting bullet points into paragraphs (34\%, shallow).

Reviewing was dominated by surface-level proofreading, such as grammar and spelling checks (75\%, shallow), though substantial minorities used AI to identify unclear sections (43\%, deep) or provide general evaluative feedback (36\%, deep), consistent with the strong bidirectional association between drafting and reviewing in Table~\ref{tab:cooccurrence}. Taken together, these patterns suggest that AI is used here primarily as a linguistic and editorial support tool, focused on polishing and improving existing text.

\paragraph{Peripheral cluster (early--late links).}
Peripheral combinations linking ideation or planning with reviewing were also common (e.g., ideation--reviewing: 48\%; planning--reviewing: 57\%), alongside moderate links between planning and drafting (55--57\%). Students in this pattern used AI to initiate work at the beginning of the process and to provide reassurance or checking near submission, without sustained engagement during intermediate stages. Stage-level data suggest this typically involved shallow ideation support (e.g., generating explanations or ideas) combined with shallow reviewing support (e.g., grammar checks or automatic improvements).

Overall, these patterns indicate that students rarely adopt AI as an end-to-end writing partner. Instead, they deploy AI selectively at specific points that align with immediate needs, such as sensemaking, text production, or entry and exit support, rather than uniformly across all stages.

\subsubsection{Comparative analysis}

This analysis examined how individual factors differed between AI users and non-users at each stage of writing. Table~\ref{tab:factors_long} summarizes these comparisons across stages and factors, reporting group means and pairwise tests for each factor at each stage. We focus here on high-level patterns rather than individual statistics.

At early stages of writing, AI use was associated with more positive evaluations of its role in supporting learning, productivity, and exploration. At ideation, users generally reported higher trust in AI and stronger perceived benefits for learning, motivation, productivity, writing quality, personal style, and text ownership than non-users. Planning showed a similar pattern: users emphasized learning, productivity, and quality gains, whereas non-users placed greater emphasis on preserving authorial identity.

At later stages, contrasts shifted toward a clearer tension between perceived performance benefits and concerns about authorship. At drafting and reviewing, users again generally reported higher trust in AI and stronger perceived benefits for writing quality and productivity, while non-users expressed greater concern for authorial identity and ownership. In contrast, sourcing showed no consistent differences between users and non-users across factors. Overall, academic writing confidence did not consistently differ between users and non-users, whereas most other factors differentiated the groups at one or more stages.

\subsection{Discussion}

This study provides empirical evidence that students’ engagement with AI writing tools is selective, task-dependent, and organized around distinct needs within the academic writing process. First, rather than adopting AI uniformly, most students concentrated their use within two to three stages of the writing process. Few applied AI across the entire process, and no single stage dominated. Co-occurrence patterns suggested that AI use typically clusters around early stages (ideation, sourcing, planning), late stages (drafting and reviewing), and certain combinations of these stages. This aligns with prior accounts of flexible, stage-based adaptation in academic writing \cite{shibani_untangling_2024, arowosegbe_perception_2024}.

Second, the factors distinguishing AI users from non-users varied systematically across stages. In the early stages (ideation and planning), AI users generally reported greater perceived benefits in productivity, learning, motivation, and writing quality. Non-users, in contrast, tended to express lower trust in AI and stronger concerns around ethics and ownership, and in some cases higher AI literacy. These findings mirror prior research, suggesting that AI's most valued affordances in early stages are idea generation and structural organization \cite{arowosegbe_perception_2024, strzelecki_use_2024, boillos_student_2025}. More literate students may also be more cautious about relying on AI due to concerns about originality and reliability \cite{johnston_student_2024}.

In the middle to later stages (drafting and reviewing), the differences focused more on authorial identity and agency. AI users associated the tool with improvements in writing quality, trust, and productivity, while non-users maintained stronger commitments to personal authorship. This suggests that concerns about authorship and personal voice become more salient in these stages. By contrast, AI was more often framed as a productivity tool in earlier stages.

Notably, what appeared to distinguish AI users from non-users was not writing confidence but how they evaluated the tool’s benefits, risks, and legitimacy. Writing confidence did not consistently differ between AI users and non-users across stages \cite{elander_complex_2006, arowosegbe_perception_2024, johnston_student_2024, boillos_student_2025}. Instead, AI adoption appeared to be more closely associated with students' perceptions of the tool's usefulness, trustworthiness, and ethical acceptability. This suggests that engagement with AI is shaped by value judgments concerning its benefits, risks, and responsibilities, rather than a compensatory strategy for writing deficiencies.

Stage differences were evident not only in whether students used AI, but in how they engaged with it. Deeper, more reflective engagement was most common during drafting and planning, while shallower, procedural use was more frequent in ideation, sourcing, and reviewing. This aligns with research suggesting that central stages of writing—focused on argumentation and coherence—require greater metacognitive control, while earlier and later phases are more conducive to discrete tasks and surface-level assistance \cite{yao_qualitative_2025, parker_negotiating_2024, shibani_untangling_2024}.

Finally, the background results align with broader survey trends: students reported frequent AI use across a variety of assessments, with ChatGPT being the most commonly used tool \cite{freeman_student_2025, arowosegbe_perception_2024}. Participants in this study demonstrated above-average AI literacy and writing confidence, indicating a cohort that was critically engaged with the tool, rather than relying on AI to compensate for perceived weaknesses. However, disclosure comfort varied, reflecting persistent concerns over transparency and institutional ambivalence toward AI use \cite{luo_jess_critical_2024, perkins_artificial_2024, boillos_student_2025, xia_social_2026}.

\subsubsection{Limitations}

These findings provide a structured overview of how AI use is distributed across writing stages and how it relates to individual factors. However, they are constrained by several aspects of the study design. Although appropriate for an exploratory study, the reliance on single \textit{ad hoc} items limits the robustness and interpretability of the measured constructs. In addition, sample size and power constraints restrict the analysis of more complex relationships between factors. Accordingly, the results should be interpreted as indicative patterns rather than definitive accounts of how these factors interact. Future work could build on this by employing validated multi-item measures and larger samples to enable more robust modeling of these relationships.

Beyond these design-specific considerations, survey-based approaches introduce further limitations. Self-report measures provide limited control over how participants interpret constructs such as “use,” “trust,” or “authorship,” and may not capture the contextual reasoning underlying their responses. These challenges are particularly salient in the context of AI use in education, where reported use may underestimate actual behavior due to social desirability bias and selective disclosure \cite{ling_underreporting_2026}, as well as reputational risks and social dynamics surrounding AI use, where individuals may intentionally conceal or obscure their use to signal competence or avoid negative judgment \cite{sarkar_ai_2025, xia_social_2026}. Prior HCI research further suggests that disclosures are not simple reflections of behavior, but are shaped by trust, perceived credibility, and the relational dynamics between researcher and participant \cite{ruitenburg_what_2026}. In this context, self-report data may not only under-represent actual behavior but systematically reflect the social conditions under which participants feel able to disclose it.

To address these limitations, a second study was conducted employing in-depth interviews to elicit more open and contextually grounded accounts of AI use in practice. Unlike fixed-response formats, interviews allow for clarification, probing, and exploration of the reasoning underlying participants' decisions. The study focuses on students who were both trained in and permitted to use AI, enabling more informed and candid reflection on their practices. By building rapport and shared ground within the interview interaction, the design aims to mitigate the disclosure constraints identified above and capture how AI integration is negotiated across the writing process. Study~2 thus complements Study~1 by moving from the distributional patterns identified in the survey to the enacted workflows and decision-making processes that underlie them. We now turn to this study.

\section{Study 2: Interview Study of Personalized AI Integration Patterns}

\subsection{Methods}

\subsubsection{Participants}

Twelve participants were recruited through convenience sampling from a postgraduate HCI program that provided explicit training in critical AI integration and formally permitted assistive use of AI. This context enabled participants to discuss their AI practices openly and ensured familiarity with adapting such tools within their own writing workflows. It also reduced some of the disclosure constraints often associated with AI use in academic settings, making the sample well suited to examining how students reflect on and articulate established integration practices. Inclusion criteria required prior use of AI in academic assessment, ensuring alignment with the study’s focus on established and personalized integration practices. Each participant received £10 compensation, and the study was covered by the same ethical approval as Study~1.

\subsubsection{Materials and Procedure}

Prior to the interviews, participants completed a brief Qualtrics questionnaire to provide informed consent, verify inclusion criteria, and characterize the sample. After reviewing the information sheet, they confirmed prior AI use in written assessments and completed the same 5-point Likert measures as in Study~1: AI literacy (Chan and Zhou’s 7-item AI Knowledge subscale) \cite{chan_expectancy_2023}, academic writing confidence (three \textit{ad hoc} items assessing perceived ability, feedback received, and ability relative to peers), and comfort disclosing AI use (three \textit{ad hoc} items assessing disclosure in written work, to lecturers, and to peers).

Semi-structured interviews were conducted using a guide designed to support systematic yet flexible exploration, allowing for follow-up probes and elaboration. Interviews began with a discussion of participants’ academic writing practices and experiences with AI use, followed by a brief visual probe based on Shibani et al.’s writing-stage model \cite{shibani_untangling_2024} to elicit workflow-level overviews.

The critical incident technique \cite{flanagan_critical_1954} was then used to elicit detailed accounts of AI use across writing stages. Participants were asked to recall a recent assessed assignment, anchoring their reflections in concrete experience and enabling consideration of how AI use influenced outcomes. This approach supported the reconstruction of situated decisions across the writing process rather than eliciting only general attitudes toward AI use.

Interviews followed the guide as a flexible framework, ensuring comprehensive yet adaptive coverage. Sessions lasted 60–90 minutes; six were conducted in person on campus and six online via Microsoft Teams, based on participant preference. All interviews were audio-recorded and subsequently transcribed for analysis.

\subsubsection{Data Analysis}

Responses from the background questionnaire were summarized using descriptive statistics in R. Interview transcripts were anonymized and analyzed in NVivo following Braun and Clarke’s reflexive thematic analysis \cite{braun_using_2006}, enabling iterative theme development while maintaining interpretive reflexivity. Researcher positionality as a fellow student and AI user was addressed through reflexive notes capturing how rapport, shared context, and resonance with participants shaped interpretation, in line with established qualitative guidelines \cite{braun_using_2006, olmos-vega_practical_2023} and recent work highlighting the role of shared identity in participant–interviewer dynamics \cite{ruitenburg_what_2026}. Both semantic and latent coding were used to capture participants’ explicit accounts and underlying rationales. An iterative combination of inductive and deductive approaches balanced data-driven insights with theoretical alignment to prior literature and Study~1 findings.

Initial inductive coding identified AI use practices and motivations without imposing pre-existing categories. Specific actions (e.g., requesting ideas, assessing novelty) were grouped into broader categories such as brainstorming support or idea evaluation, while motivational codes (e.g., saving time, plagiarism concerns) were clustered into orientations such as efficiency focus or originality concerns.

In a subsequent deductive phase, these categories were reclassified using Shibani et al.’s CIAW framework \cite{shibani_untangling_2024} and survey factors to ensure conceptual continuity across studies. Practices were categorized as deep (iterative and critical engagement), shallow (limited evaluation of AI outputs and increased passive reliance), or no use (explicit resistance) \cite{shibani_untangling_2024}. Motivational orientations were aligned with survey constructs; for example, efficiency focus mapped onto productivity, while originality concerns reflected ownership.

Individual interaction profiles summarizing each participant’s stage-based use and attitudes were then synthesized into an integrated thematic framework. The themes were developed iteratively through cross-participant comparison, supported by visual mapping of individual usage patterns across stages to identify recurring configurations of behavior. Following Braun and Clarke’s emphasis on analyzing patterns across the dataset rather than isolated codes \cite{braun_using_2006}, the final framework comprised three primary themes capturing distinct patterns of AI-supported writing workflows.

\subsection{Results}

Participants scored higher than the Study~1 sample in openness to disclose AI use ($M$ = 4.03, $SD$ = 1.14), AI literacy ($M$ = 4.77, $SD$ = 0.56), and academic writing confidence ($M$ = 4.00, $SD$ = 1.06), indicating a highly proficient group well positioned to reflect on and articulate their practices.

Three core themes were constructed to capture how AI is integrated into writing. These are defined by differing trajectories of reliance across writing stages and their underlying motivations. The themes are intended as illustrative profiles rather than fixed categories, as most participants exhibited elements of multiple themes, while aligning more closely with one dominant pattern. The following sections elaborate these themes through their stage-specific practices and associated motivations.

\subsubsection{Quality-Oriented: Focus on Enhancing Performance and Managing Risk}

Quality-Oriented students consistently prioritized the quality of their final product, showing limited tolerance for efficiency trade-offs and a conservative approach to sharing agency with AI. As one participant explained: “\textit{I know it can give me something fast, but I don’t trust it to be good and get me a good grade unless I’ve gone through it myself}” [P11]. Their use patterns typically shifted from independence in earlier stages to more iterative integration in later stages, reflecting what Shibani et al. describe as movement between no or little use and deeper forms of engagement \cite{shibani_untangling_2024}. These participants demonstrated high awareness of AI’s limitations, particularly the risk of misleading outputs, as noted by P6: “\textit{AI can convincingly sound right but actually be very wrong}.”

This orientation reflects a form of risk-sensitive engagement in which students weigh potential quality gains against concerns about accuracy, reliability, and ownership. Prior work similarly shows that trust in AI, perceived output quality, and AI literacy shape when and how students incorporate AI into writing, often leading to selective and closely supervised use \cite{chan_expectancy_2023}.

\paragraph{Minimal use in Ideation and Sourcing to protect originality and accuracy.}
At the start of the process, performance-driven participants favored independent ideation and sourcing to ensure arguments were grounded in literature, protecting text quality. This aligns with prior accounts of concerns around generic outputs and misleading sourcing as barriers to open use \cite{shibani_untangling_2024}, with broader survey evidence highlighting concerns about reliability and appropriate use \cite{johnston_student_2024}. Participants explicitly described these risks: “When it generates ideas for you from scratch, ChatGPT tends to be generic or a basic” [P11]; “\textit{If you start completely blank, it’s really easy to get influenced by whatever the first thing the AI gives you… you end up ideating just with ChatGPT and picking very banal topics}” [P9]; “\textit{AI can hallucinate sources…or give you the summary of another paper when you enter the title of a specific article}” [P6]; and “\textit{AI is not good at summarizing long-form paper content accurately}” [P11].

AI was therefore restricted to preliminary brainstorming, with outputs substantially criticized or reshaped: “\textit{I use it for ideation sometimes but I tend not to use the ideas it gives me as they are, I go through multiple rounds of very basic ideas and then put them together through my own insight}” [P9]. A similar pattern appeared in sourcing, where AI was treated as a general overview tool rather than a substitute for scholarly search: “\textit{AI may give you the groundwork, like the foundational papers, but when it comes to more niche and specific ones… not so much. I’ll still go on Google Scholar and find my own sources}” [P11]. For these students, originality, accuracy, and writing quality were inseparable, shaping a cautious stance on ideation and sourcing.

\paragraph{Minimal use in Planning to preserve argument rigor.}
Planning was described as the phase where ideas and sources were deliberately integrated into a coherent argument, with AI often deliberately avoided. As one participant stated: “\textit{I’ll manually write a skeleton…I don’t want AI to spit an outline out for me}” [P1]. Another reinforced this: “\textit{The structure is where I make sense of the sources and shape the argument, so I prefer to do it independently}” [P11].

Planning was therefore constructed as a meaning-making phase where ownership was essential to performance. As P6 put it: “\textit{I think [when planning] it’s really important to be in that muddy area of getting dirty and being stuck because I think it’s so important to helping you like get a better understanding of the nuances and what you’re trying to talk about and structure it better}”. This resonates with prior descriptions of planning as a central site of authorial control and learning, where automated support can be perceived as compromising rather than enabling quality \cite{shibani_untangling_2024}.

\paragraph{Deep use in Drafting and Reviewing to refine expression and clarity.}
In later stages, performance-driven students shifted from avoidance to deep, iterative use. Drafting and reviewing were viewed as phases where AI could reliably add value by polishing language and improving clarity rather than generating arguments from scratch. As participants explained: “\textit{In terms of language and conciseness, AI does a lot}” [P11]; “\textit{I’ll polish the tone and the structure…and then sometimes I’m over the word count, so I’ll use AI to reduce it, or to make something clearer}” [P6].

Writing background also shaped this reliance, with one participant noting: “\textit{I did computer science and before university my A levels were just STEM, so I have not had a lot of experience writing long pieces of writing}” [P6]. This reflects prior accounts of AI’s potential benefits in language refinement and editing, especially for students with lower confidence in expression \cite{kim_exploring_2025}.

Even in these later phases, participants emphasized retaining narrative control: “\textit{It helps me write faster, but I’m still controlling the whole narrative}” [P5]. Authorship concerns were muted here, with P9 stating: “\textit{Academic writing is quite objective, so if it’s reliable it doesn’t really matter who wrote the specific wording in the end}”. Drafting and reviewing were thus framed as polishing stages, where students leveraged AI to refine expression and clarity while maintaining control over argumentation and structure.

\subsubsection{Learning-Oriented Use: Focus on Broadening Exploration and Protecting Authorship}

Learning-oriented participants described using AI in a pattern virtually opposite to quality-oriented users, with heavier use in the early stages to support exploration, followed by progressive discontinuation as writing advanced. Early phases often involved iterative prompting and critical comparison, mapping onto what Shibani et al. describe as deep use along the continuum of interaction \cite{shibani_untangling_2024}, while drafting and reviewing typically reflected no use or tightly constrained assistance.

Students in this group generally expressed higher trust in AI as a learning scaffold, as stated by P8: “\textit{I would say AI gave me a lot of general knowledge}”. Yet, they were equally concerned about deskilling and authorship boundaries, as shown in statements such as: “\textit{I fear my skills will atrophy (with AI use), even with using it for minor editing I may lose the patience to sit there for a really long time to find the perfect way to put this sentence together}” [P10]. These motivations align with social constructivist accounts of AI framed as a potential learning scaffold that simultaneously risks undermining academic integrity and ownership when over-relied upon \cite{boillos_student_2025, song_enhancing_2023, vygotsky_mind_1978}.

\paragraph{Deep use in Ideation and Sourcing to expand understanding.}
At ideation and sourcing, AI was used to widen the space of possibilities, generate starting points, and support comprehension of dense material. As P1 explained, “\textit{It helps me to find information that supports or maybe sometimes contradicts a certain narrative I already have in my head from the brief}”. Similarly, AI was described as a “\textit{librarian}” for “\textit{the direction of information finding}” [P1]. For sourcing, it often served as a pointer rather than an authority, and participants used it iteratively to grasp complex concepts, as P4 noted: “\textit{Not coming from a science background, for materials that are quite dense it’s definitely helpful for your learning to have something ‘dumbed down’}”.

Despite this openness, students critically evaluated outputs, highlighting their retained control: “\textit{Sometimes it does give me something that’s not too related… but I think that’s fine because I still have the control. I can make my own judgments}” [P7]; “\textit{I make sure to put AI suggested information through a search engine to triangulate that to see if other sources say similar things}” [P1]. This exploratory orientation set the groundwork for planning, where AI could provide structural scaffolding while students retained ownership of interpretive work.

\paragraph{Deep use in Planning to provide structural support.}
Learning-oriented participants were more willing than quality-oriented peers to allow limited AI input when planning. For instance, P7 admitted: “\textit{I use AI to structure the sections of papers because I am not that used to that}”, while P8 emphasized: “\textit{It’s good for giving a skeleton, but I still keep my own tone and organization}”.

For others, AI input was restricted to less critical components, as P4 reflected: “\textit{The introduction was probably inspired heavily by ChatGPT because it’s not the important part… just a summary}”.

Planning was thus treated as a space where AI could assist with structure and formulaic tasks, while argumentative logic and interpretation remained the student’s domain. This pattern carried forward into drafting and reviewing, where boundaries around authorship and identity became more rigid.

\paragraph{Minimal use in Drafting and Reviewing to preserve ownership and authorial identity.}
In drafting, students consistently marked interpretation and voice as non-delegable. P4 explained: “\textit{I’ll use it to summarize, but I don’t use it for analysis, that part has to be mine}”. P12 reinforced this: “\textit{I tend to avoid…writing the interpretation bits…because that’s the part where my personality and identity get into the text}”.

Writing itself was framed as integral to understanding, as P1 stressed: “\textit{Writing itself is how I clarify my argument, so if I don’t do that part, I don’t really understand what I’m saying}”. For others, authorship was tied to identity and pride: “\textit{I feel like academic writing is part of who I am as a student, so I wouldn’t want AI to take that away}” [P8]; “\textit{I worked in writing, I am a writer, I would be ashamed to call myself one if I let AI write for me}” [P10]. Concerns also extended to skill retention: “\textit{I write well, and I don’t want to lose that ability while others just let AI write for them}” [P10].

In reviewing, AI was often dismissed as unhelpful or too generic: “\textit{It doesn’t have a sense of judgment, it just tells you everything is fine}” [P3]; “\textit{At the end of the day a human judge will read it, so I prefer to trust myself}” [P7]. Some saw polishing as potentially undermining coherence: “\textit{It’s not good at storytelling… it might make sentences smoother, but it loses the flow of the argument}” [P7].

\subsubsection{Productivity-Oriented Use: Focus on Efficiency and Sustaining Momentum}

Productivity-oriented students allowed the most reliance on AI across the workflow, showing fewer deliberate points of discontinuation than performance- and learning-focused peers. Their stance was framed less around fine-grained task fit or personal values, and more around maintaining momentum and reducing friction in the writing process. Participants described using AI to alleviate overwhelm, keep moving forward, and lower the activation energy of writing, often granting the tool more agency than students in other profiles, as described by P2: “\textit{I treat AI like a junior analyst, or a junior me}.”

Students in this profile frequently highlighted AI’s motivational role: “\textit{Sometimes it just gets me started… once I see something on the page, I can keep going}” [P2]. This orientation aligns with prior research indicating that students value AI for its ability to support productivity, lower cognitive effort, and sustain engagement during writing \cite{dhillon_shaping_2024, fernandes_performance_2026}. From this perspective, AI serves as a form of cognitive offloading and scaffolding rather than a tool primarily assessed based on output quality or authorship, helping students maintain progress even when intrinsic motivation is low.

\paragraph{Shallow use in Ideation and Sourcing to enhance efficiency.}
During ideation and sourcing, productivity-oriented students relied heavily on AI to generate starting points and frameworks. Unlike performance- and learning-oriented peers who approached these phases more conservatively, these participants were comfortable using AI extensively to structure directions and summarize sources. As P2 noted, “\textit{I’ll ask it to give me some ideas or starting points… even if I don’t keep them, it means I’m not staring at a blank page}” [P2].

This willingness did not mean they ignored AI’s limits, but they were prepared to make trade-offs in quality or ownership for efficiency gains. As P5 explained, “\textit{It gives me a direction… sometimes it’s wrong, but it saves me time, and I can just fix it}” [P5].

\paragraph{Deep use in Planning and Drafting to sustain momentum.}
In planning and drafting, productivity-oriented students described using AI in an iterative yet strongly goal-oriented way. They often delegated paragraph initiation or outline drafting to the tool, then refined outputs with their own edits. P2 explained, “\textit{I’ll get it to write a paragraph from my notes… then I’ll edit it down. It just speeds things up}” [P2]. For P5, AI was particularly useful for structuring arguments quickly: “\textit{I give it a bullet list, it gives me a draft paragraph, and then I make it sound right}” [P5].

Although this mirrored the iterative use seen in the other two profiles, the emphasis was different: where quality-oriented students used iteration to safeguard quality, and learning-oriented students used it to scaffold understanding, productivity-oriented students used it primarily to maintain momentum and meet immediate writing goals. Their accounts reflect a pragmatic orientation toward keeping work moving, with less attention to optimizing task fit or preserving strict boundaries around authorship.

\paragraph{Deep use in Reviewing to accelerate checks and ensure submission suitability.}
Reviewing was the phase where productivity-oriented students showed the strongest reliance on AI, often describing it as a critical reassurance tool. They demonstrated sophisticated prompting strategies for feedback and revision, showing trust in the tool’s capacity to highlight weaknesses. P5 described routinely submitting drafts for a holistic check: “\textit{Each paragraph has already been through AI, and then I’ll run the whole essay through again to see what it says}” [P5]. P2 highlighted the motivational aspect of this process: “\textit{I like to ask it to critique my work… even if I don’t take everything, it makes me feel more confident to hand it in}” [P2].

While they remained somewhat critical in incorporating AI’s feedback, these participants were clearly more trusting than other groups. As P4 explained, “\textit{I’ll get it to mark against the rubric… it points things out that I wouldn’t have noticed, and that’s useful}” [P4]. For productivity-oriented students, reviewing was less about preserving integrity than about maximizing productivity and assurance. Their reliance at this stage underscores how motivations of momentum and reassurance shaped use across the workflow, in contrast to the quality-oriented stance of performance-focused peers and the principled boundaries of learning-oriented ones.

\subsection{Discussion}

Findings from Study~2 corroborate the survey patterns and provide a richer account of how stage-based AI use is assembled into coherent writing workflows. Rather than reflecting stable “types” of users, the three profiles show how students organize AI use around competing priorities, selectively allocating control and effort across the writing process.

Quality-oriented students, primarily concerned with final product quality, showed greater skepticism at ideation and sourcing, citing low trust, concerns about reliability, and potential performance impairment. They nonetheless accepted linguistic assistance at drafting and reviewing as a pragmatic trade-off. Their high AI literacy supported informed decisions about where AI could enhance quality without compromising control \cite{kim_exploring_2025, kim_students-generative_2025}. For these students, authorship resided in conceptual decision-making, while reduced control over phrasing was tolerated in exchange for clarity and performance. This reflects Dhillon et~al.’s “ghostwriter effect” \cite{dhillon_shaping_2024}, where perceived quality increases despite diminished authorship when users retain oversight of outputs. Retaining early-stage control thus enabled quality gains, in contrast to more passive or linear patterns of AI use associated with weaker engagement and lower-quality outcomes \cite{nguyen_human_ai_2024}. Overall, this profile reflects a performance- and risk-oriented strategy in which AI is evaluated not in terms of convenience, but in terms of its contribution to output quality and reliability.

Learning-oriented students used AI as a scaffold in early stages but resisted it during drafting and reviewing, where they prioritized personal expression, narrative coherence, and skill development. Their boundaries reflected principled views on authorship and concerns about deskilling or loss of voice. Writing itself was treated as integral to understanding, consistent with evidence that composition supports cognitive processing \cite{elander_complex_2006}. Their reluctance to delegate textual production parallels attitudes among creative writers who emphasize authenticity and voice \cite{hwang_it_2024, guo_pen_2025, varanasi_ai_2025}. This pattern aligns with accounts where early support is used to broaden exploration but withdrawn as students move toward ownership and expression \cite{boillos_student_2025, song_enhancing_2023}. Here, AI is integrated selectively as a learning aid, with clear boundaries that preserve authorship.

Productivity-oriented students exhibited the broadest reliance on AI, using it across both early and late stages to reduce friction, maintain momentum, and gain reassurance. Their approach prioritized usefulness and motivation over performance optimization, emphasizing AI’s role in lowering cognitive effort and supporting self-efficacy. This willingness to trade some quality for efficiency reflects a pragmatic orientation toward progress and completion, rather than the optimization of output \cite{dhillon_shaping_2024, fernandes_performance_2026}. 

Similar patterns have been observed in prior work, which highlights AI’s role in reducing writing apprehension and sustaining engagement \cite{boillos_student_2025}. Importantly, this reliance reflects a deliberate prioritization of progress rather than a lack of awareness. 
Consistent with cognitive offloading perspectives, AI is viewed as an external resource that redistributes effort and sustains workflow continuity \cite{sweller_cognitive_1988}. Recent studies also suggest that higher confidence in AI use correlates with reduced cognitive effort, where students focus more on overseeing the process rather than actively engaging in writing tasks \cite{LeeCHI2025}.

Overall, these findings show that student AI use cannot be reduced to binary distinctions between critical and non-critical practice. Instead, AI-supported writing is organized around value-based trade-offs among learning, quality, productivity, and authorship, with students allocating control differently across stages of the writing process. This supports viewing AI integration as a form of self-regulated learning, in which strategies are adapted to individual goals and constraints, consistent with models of self- and socially regulated learning \cite{nguyen_human_ai_2024, nguyen_hybrid_2024, jarvela_predicting_2023}. Extending stage-based models of critical adaptation \cite{shibani_untangling_2024}, these results suggest that “critical use” is not a stable trait, but a context-sensitive practice shaped by how students resolve competing priorities within their workflows.

\subsubsection{Limitations}

While Study 2 was designed in part to address the limitations of self-report survey data, it introduced its own constraints that should be considered. The sample was well-suited to examining established AI integration practices. However, it was also limited in size and composed of students with high AI literacy and writing confidence, which restricts the generalisability of the workflow profiles identified. Participants exhibited ceiling effects for AI literacy and writing confidence relative to the broader survey sample, and deep engagement strategies --- iterative prompting, critical evaluation, and selective incorporation of outputs --- predominated throughout, with shallow use largely confined to sourcing. This likely reflects the effects of explicit AI literacy training within a supportive institutional context, consistent with prior evidence that structured instruction enables more critical and reflective AI use~\cite{luo_jess_critical_2024, perkins_artificial_2024}. The profiles identified here may therefore over-represent deliberate, well-regulated forms of integration, and different configurations may emerge in populations with lower familiarity or less structured guidance. With only 12 participants, the sample is also small, and the profiles should be treated as illustrative rather than exhaustive.

The interpretation of these findings should also consider issues of disclosure and researcher positionality. As noted in Study~1, self-reported AI use reflects not only behavior but also willingness to disclose, particularly under conditions of policy ambiguity or perceived risk \cite{ling_underreporting_2026}. In this study, the explicit permissibility of AI use and the use of in-depth interviews likely supported more open, contextually grounded accounts \cite{ruitenburg_what_2026}. The interviewer’s peer status may have further reduced perceived evaluative pressure, consistent with evidence that students are more willing to disclose AI use to peers than to institutional authorities \cite{ling_underreporting_2026}. However, accounts should still be understood as situated, as the formal setting of a research study likely shaped how practices were articulated. These findings therefore reflect both participants’ practices and the conditions under which they were reported, rather than providing a complete representation of underlying behavior.

\section{General Discussion}

The research reported here shows that students’ use of AI in academic writing, a central form of knowledge work developed during university education, is structured, selective, and embedded within the organization of writing workflows, rather than applied uniformly across the writing process. 

Study~1 provides a systematic mapping of these patterns. Students typically engaged AI in only two or three stages, forming three recurring clusters: an early-stage cluster (ideation, sourcing, and planning), a late-stage cluster (drafting and reviewing), and a peripheral cluster linking ideation and reviewing. These clusters were associated with distinct configurations of perceived benefits and concerns. Learning-related benefits were most salient at the start of the process; quality gains and authorship concerns converged during drafting and reviewing; and productivity benefits appeared at both entry and exit points, supporting task initiation and completion. Depth of engagement followed a similar structure, with deeper, iterative use concentrated in planning and drafting, and more procedural use in sourcing and reviewing. Taken together, these patterns show that AI integration reflects shifting trade-offs between learning, productivity, quality, and authorship across the writing process.

Study~2 extends this account by showing how these stage-based patterns are assembled into coherent, end-to-end workflows by critically engaged students. Three recurring, value-oriented profiles were identified. \textit{Learning-Oriented} students concentrated AI use in early stages to support exploration and understanding, withdrawing later to preserve authorship and learning. \textit{Quality-Oriented} students limited early use but adopted deeper, iterative integration during drafting and reviewing to enhance clarity and performance. \textit{Productivity-Oriented} students used AI more broadly, with peaks at the beginning and end of the process, prioritizing efficiency, momentum, and reassurance. These profiles demonstrate how stage-level patterns, factor associations, and depth of engagement combine into coherent strategies, showing that AI use emerges from situated trade-offs negotiated differently across workflows.

Across both studies, AI use is not best understood as a general tendency to adopt or reject tools, but as a pattern of stage-specific integration shaped by shifting priorities. Taken together, the findings reshape AI use in academic writing from a question of adoption to one of allocation, suggesting how control, effort, and responsibility may be distributed across stages of the writing process. The following sections develop the conceptual, methodological, and practical implications of this shift.

\subsection{Conceptual and Methodological Implications}

The patterns identified in this work suggest that AI-supported writing cannot be fully captured by single-lens theoretical accounts. Rather than reflecting a uniform mode of “successful” use, students organize AI integration around competing priorities such as learning, quality, productivity, and authorship. These orientations align with various theoretical perspectives. On the one hand, constructivist learning and scaffolding perspectives \cite{ou_academic_2024, parker_negotiating_2024, vygotsky_mind_1978} emphasize the role of AI in supporting the learning process by providing resources for active knowledge construction. On the other hand, cognitive offloading frameworks \cite{sweller_cognitive_1988, dhillon_shaping_2024, fernandes_performance_2026} highlight AI’s role in reducing cognitive load and redistributing effort to maintain productivity and mental resources. However, no single framework adequately captures how these competing priorities coexist and inform real-world writing workflows. This points to the need for an integrative perspective in which AI-supported writing is understood as a problem of allocation rather than adoption, focused on how users distribute control, effort, and responsibility across stages of work. This aligns with broader perspectives on human–AI co-creation, where interaction dynamics such as turn-taking, contribution, and communication structure the collaborative process and shape outcomes \cite{rezwana_designing_2023}.

These findings also have important implications for research design. The stage-specific and selective ways in which participants delegated work to AI are unlikely to be fully captured in tightly controlled tasks that fix the point of intervention or constrain interaction patterns. While such designs can isolate mechanisms, they risk obscuring how AI use is assembled and revised across the writing process. To better capture these dynamics, future work should adopt more naturalistic and longitudinal approaches that examine how decisions about delegation evolve over time and across stages, while retaining sufficient structure to enable comparison across contexts.

\subsection{Practical Implications for Education, Assessment, and Tool Design}

The current findings have practical implications for how AI use is supported in education, how writing is assessed, and how AI writing tools are designed. From an educational perspective, our results suggest that approaches centered on prohibition or general awareness of AI’s risks are insufficient for preparing students to navigate the practical decisions that arise across a writing workflow, consistent with prior literature \cite{perkins_artificial_2024}. Effective training should instead help students understand where AI can support exploration and sense-making, where it risks undermining authorship or learning, and how to retain control over core intellectual contributions. Rather than offering only generic guidance, such training should be adaptive, helping students identify what value AI can offer at each stage of the writing process and where it can be integrated without compromising their intellectual work.

A related implication concerns assessment design. The findings show that students integrate AI selectively across the writing process, using it in different ways depending on the stage, the value they assign to the tool, and the extent to which they feel authorship must remain their own. From this perspective, the central question is not simply whether students use AI, but what forms of engagement an assessment is intended to elicit and evaluate. Responding to AI uptake in university settings therefore requires educators to be explicit about what a task is actually measuring, and which writing or broader sense-making activities are essential to demonstrating competence. Only on that basis can boundaries around AI use be drawn meaningfully rather than applied uniformly. If the aim is to evaluate what students actually understand, then assessment must create opportunities for that understanding to be demonstrated directly, whether through sampled oral examination, such as viva-style assessment, or through unseen written exams where appropriate. Such approaches can help make understanding and authorship more visible, even where AI-assisted writing is permitted.

Finally, the findings have important implications for the design of education-oriented AI writing tools. Much existing work focuses on supporting drafting and revision through iterative interaction and stylistic refinement \cite{suh_luminate_2024, reza_abscribe_2024, lin_rambler_2024}. However, the results presented here indicate that earlier stages of writing, such as ideation and planning, are equally critical points of integration, where students balance efficiency with depth of engagement. Tools should therefore support stage-aware interaction, enabling users to adjust levels of automation, maintain transparency over system contributions, and engage in iterative and reflective workflows across the entire writing process, consistent with prior work emphasizing the role of interaction design in shaping human–AI co-creative partnerships \cite{rezwana_designing_2023}. Importantly, different users may require different forms of support. Those with strong authorship boundaries may resist generative assistance in drafting, while others may prioritize tools that reduce friction and sustain momentum. Designing for this diversity requires moving beyond one-size-fits-all solutions toward more flexible and adaptive systems.

\subsection{Limitations and Future Work}

Several limitations should be considered when interpreting these findings. First, the samples used in both studies may not fully represent the broader student population. The survey sample, while sufficient to detect broad patterns, was relatively small, and the interview sample relied on convenience recruitment within a context where AI use was explicitly supported. Participants in Study~2 also exhibited relatively high levels of AI literacy and writing confidence, which may have led to an over-representation of more reflective and deliberate forms of engagement. As a result, the workflow patterns identified here may differ from those of less experienced or less confident users. Future research should therefore examine whether similar stage-based patterns and workflow configurations emerge across more diverse populations, disciplines, and institutional contexts \cite{persson_we_2021, stohr_perceptions_2024}.

Second, the study relies in part on simplified measures of complex constructs. In the survey, factors such as authorship, trust, and learning were captured using single-item \textit{ad hoc} measures, while interviews addressed these concepts through open-ended discussion. Although appropriate for exploratory analysis, this limits the precision and comparability of these constructs. Future work could build on these findings using validated multi-item scales \cite{cheung_development_2017, mitchell_development_2021} and larger samples to support more robust modeling of relationships between individual factors and patterns of AI use.

Third, both studies rely on self-reported accounts of AI use, which are subject to social desirability bias and selective disclosure. As recent work has shown \cite{ling_underreporting_2026}, students may systematically under-report AI use, particularly in contexts where policies are unclear or perceived as restrictive. More broadly, disclosure in research settings is shaped by interactional context and participants’ judgments about what is appropriate to share \cite{ruitenburg_what_2026}. While the interview design aimed to mitigate this by creating conditions for open reflection, responses should still be understood as situated accounts shaped by context and perceived expectations.

Finally, the study focuses primarily on student perspectives, whereas AI-supported writing is embedded within a broader ecosystem involving educators, institutions, and tool designers \cite{luo_jess_critical_2024, perkins_artificial_2024}. Understanding how AI is integrated into academic writing therefore requires attention to these multiple perspectives. Future research could extend this work through collaborative and participatory approaches, such as co-design, to examine how different stakeholders shape and respond to AI-supported writing practices.

\section{Conclusion}

This research shows that AI integration in academic writing is neither uniform nor reducible to binary notions of “critical” and “non-critical” use. Instead, students organize AI use in stage-specific ways, selectively integrating it across writing workflows in response to competing priorities such as learning, quality, productivity, and authorship. These findings reframe AI use in academic writing from a question of adoption to one of allocation: how control, effort, and responsibility are distributed across the writing process. Because academic writing remains a central form of knowledge work developed through university education, this shift has important implications for how writing is taught, supported, and assessed. Rather than enforcing uniform models of use, future frameworks, tools, and policies should recognize this variability and focus on supporting informed, context-sensitive integration. As AI makes text production easier, the central challenge becomes not whether students can generate writing, but how they engage with, evaluate, and take responsibility for it.

\begin{acks}
This work was carried out as part of the first author's MSc in Human–Computer Interaction at University College London, based on her MSc dissertation. The first author is now a PhD student at the University of Cambridge. We thank the participants for their time and contributions. We also thank the anonymous reviewers for their constructive feedback, which improved the clarity and presentation of this paper.

\textbf{AI Use Statement.} Generative AI tools were used in an assistive capacity to improve clarity, grammar, and readability. All content was written, reviewed, and verified by the authors. No generative AI system was used to generate analyses or results. The authors take full responsibility for the final content.
\end{acks}

\bibliographystyle{ACM-Reference-Format}
\bibliography{references}

\end{document}